\documentclass[twocolumn,showpacs,preprintnumbers,superscriptaddress]{revtex4}     
\usepackage[T1]{fontenc}  
\usepackage{amsfonts}    
\usepackage{amsmath,amsbsy,amssymb,graphicx}         
\usepackage{times}
\begin{document}                 
\title{Green's Function of Magnetic Topological Insulator in Gradient Expansion Approach}
\author{Yusuke~Hama}
\affiliation{RIKEN Center for Emergent Matter Science (CEMS), Wako, Saitama 351-0198, Japan}
\author{Naoto~Nagaosa}
\affiliation{RIKEN Center for Emergent Matter Science (CEMS), Wako, Saitama 351-0198, Japan} 
\affiliation{Department of Applied Physics, University of Tokyo, Tokyo 113-8656, Japan}
\date{\today}
\begin{abstract}{
 We  study the  Keldysh Green's function of the Weyl-fermion surface state of the 
three-dimensional  topological insulator coupled with a space-time dependent magnetization in the  gradient expansion.
Based on it we analyze the electric charge and current densities as well as the energy density and current   induced by spatially and temporally slowly-varying magnetization fields. 
We show that all the above quantities except the energy current are  generated by the emergent electromagnetic fields.
The energy current emerges as the circular current reflecting the spatial modulation of an  induced gap of the Weyl fermion. }
  
\end{abstract}

\maketitle

\section{Introduction}\label{Intro}
Topological states of matter is one of the most important topics in the modern physics.
Originating from prototypes  such as integer quantum Hall systems \cite{PrangeGirvin,DasSarmaetal1,Ezawa:2013ae} and polyacetylene \cite{polyacetylenereview}, 
 new types of topological states of matter called the topological insulator (TI) and the topological superconductor (TSC) have been discovered 
 and attracted much attention \cite{HansanKane,Qietal1,Shen,FranzMolenkamp,BernevigHuges}.
 One of the most remarkable features of the TI and  TSC is the bulk-edge correspondence.
The TI and  TSC have topologically nontrivial bulk band structure  associated with   finite topological invariants,
 while the vacuum is topologically trivial  with zero topological invariant. 
 They cannot be adiabatically connected to each other, and in order to realize the topological phase transition from the 
  topologically nontrivial sector to the topologically trivial sector, the existence of the gapless surface states are necessary.
 They are  related to the  symmetry which systems possess, for instance, the edge states in two-dimensional topological insulator are called the helical edge states
  characterized by the time-reversal symmetry, while those in the topological superconductor are called the Majorana fermions associated with the particle-hole symmetry. 
The classification of the topological states of matters by the  symmetry \cite{Schnyderetal} and finding the new topological materials are  interesting future directions.
  
\begin{figure}[t]
\begin{center}    
\centering
\includegraphics[width=0.25\textwidth]{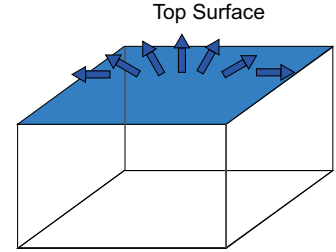}
\end{center}
\caption{The schematic illustration of the three-dimensional magnetic topological insulator.
The magnetization configuration on the top surface (the blue rectangle)
is represented by the blue arrows. }
\label{magneticti} 
\end{figure}

 Another  direction is to study the transport phenomena of the  surface states in the topological matters.  
 One candidate is the three-dimensional (3D) magnetic TI which is realized,
 for  instance, by dopping the magnetic impurities such as Cr or Mn \cite{Checkelskyetal1,Changetal,
  Checkelskyetal2,Leeetal,Kouetal,Chenetal}.
 We display the schematic illustration of the 3D magnetic TI  in FIG. \ref{magneticti}.
 The energy gap of the surface state is induced due to the exchange coupling.
 The surface state of the 3D TI is effectively described as the Weyl fermion  with spin-momentum locking.
 Various functionality owing to  charge and spin of the Weyl-surface state can be expected, and indeed, 
  rich variety of the electric and magnetic phenemona  owing to the coupling between  the magnetization 
  and the Weyl fermion have been shown both theoretically and experimentally 
  \cite{Checkelskyetal1,Changetal,
  Checkelskyetal2,Leeetal,Kouetal,Chenetal,Qietal2,Qietal3,Yokoyamaetal1,Nomuraetal,Garateetal,Hurstetal,Morimotoetal,Wangetal,Taguchietal},
  and might have potential for the applications to the spintronics \cite{Chenetal,Yokoyamaetal2}.
  Therefore, the study of the transport phenomena of the Weyl-surface state of the 3D TI controlled magnetically is important for understanding the non-equilibrium physics of the topological states of matter
  as well as the spintronics applications.

 In this paper, we theoretically investigate the transport properties of the Weyl surface state in the 3D magnetic TI.
We study the electric charge and current densities as well as the energy density and the current
 driven by the magnetization fields varying slowly with respect to space and time. 
 We use the gradient expansion approach by retaining the derivative terms of the magnetization fields up to the first order,
and derive the Keldysh Green's function describing the non-equilibrium state of the Weyl fermion. 
It is the main result of this paper, and based on it we can systematically analyze the transport phenomena of the Weyl fermion generated by the magnetization fields. 
 
 Due to the exchange interaction between the magnetization fields and the Weyl fermion,
 the in-plane magnetization fields   act as vector potentials \cite{Nomuraetal}.
 Then it is natural to introduce ``emergent'' electromagnetic fields described by the above vector potential.
 We demonstrate that the various transport phenomena induced by the magnetization fields are clearly represented as responses to the emergent electromagnetic fields.
For the energy current it cannot be described in terms of the emergent electromagnetic fields. 
It is the circular current describing a spatial modulation of the energy gap of the Weyl fermion.

 This paper is organized as follows. In Sec. \ref{gradexp},  we first present the Hamiltonian and the energy spectrum of the system.
 Then we show how the in-plane magnetization fields coupled to  the Weyl fermion and act as vector potentials, and 
 introduce the emergent electromagnetic fields. 
  Next by using the gradient expansion, we derive the Keldysh Green's function up to the first-order derivative of the magnetization fields.
The transport quantities are calculated based on this Green's function.
 In Sec. \ref{sec3electricchargecurrent},  we present the analysis of the electric charge and the current densities. 
 We show that the electric charge is induced by the emergent magnetic field, whereas the electric current density is induced by the   
  emergent electric field showing the quantized anomalous Hall effect. 
 In Sec. \ref{sec4energycurrent}, we analyze  the energy density and  current. We demonstrate that 
the energy current is the circular current described in terms of the spatial derivative of the out-
 plane component of the magnetization field.
 Sec. \ref{sec5conclusion} is devoted to the conclusion of this paper.

\section{Gradient Expansion Approach }\label{gradexp}
\subsection{Hamiltonian}\label{sec2Hamiltonian}

We analyze the dynamics of the Weyl surface state of the three-dimensional topological insulators coupled with a magnetization field varying slowly with respect to the space-time.
Focusing on one of the layer, saying the top layer, the low energy effective Hamiltonian is given by 
\begin{align}
H_0(\boldsymbol{x})=\frac{\hbar v_{\text{F}}}{i} 
\left(\sigma^y \partial_x -\sigma^x \partial_y\right)+
{m}^a(\boldsymbol{x},t)\sigma^a
-\mu\boldsymbol{1}_{2\times2},
 \label{hamiltonian1}
\end{align}
where $v_{\text{F}}$, $\mu$, and $m^a(\boldsymbol{x},t)$  $(a=x,y,z)$ are the Fermi velocity, the chemical potential,
  magnetization fields, $\sigma^a(\boldsymbol{x},t)$ $(a=x,y,z)$ are the Pauli matrices,
  and  $\boldsymbol{1}_{2\times2}$  the $2\times2$ unit matrix, respectively, with $\boldsymbol{x}=(x,y)$ being the two-dimensional coordinate. 
 Here  $m^a(\boldsymbol{x},t)$ satisfy $\sum_a (m^a)^2=m^2$, with $m$ a exchange coupling constant. The summation is taken for the repeated  index $a.$
The first term in \eqref{hamiltonian1} represents the spin-momentum locking,
while the second term ${m}^a(\boldsymbol{x},t)\sigma^a$ is the exchange term.
Here we set $\mu=0$ since we are working at  zero-temperature.

\begin{figure}[t]
\begin{center}    
\centering
\includegraphics[width=0.3\textwidth]{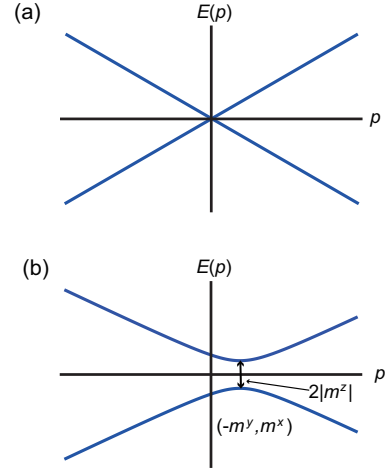}
\end{center}
\caption{The dispersion relations for the Weyl fermion. The momentum space is projected into one-dimensional space to easily see the characteristics of the dispersions.
Here we set   $v_{\text{F}}=1.$ 
(a) The dipersion relations for the Weyl fermion without the magnetization fields. 
Their dispersions are degenrate at the Weyl point (the origin).  
(b) The dipersion relations for the Weyl fermion with the magnetization fields. 
The in-plane magnetization fields shift the Weyl point to $(-m^y,m^x)$, while the out-plane magnetization field opens an energy gap with its magnitude $2|m^z|$. }
\label{weyldispersion} 
\end{figure}

The Weyl fermion has the dispersion relations
\begin{align} 
E_{\boldsymbol{p}}
=\pm\sqrt{(\check{p}^x)^2+
(\check{p}^y)^2+
(m^z)^2},    \label{dispersion}
\end{align} 
where $\check{p}^x= v_{\text{F}}p^x+m^y,
\check{p}^y=v_{\text{F}}p^y-m^x,$ with $p^{x,y}$ denoting the two-dimensional momentum. 
As presented in FIG. \ref{weyldispersion}, the in-plane magnetization fields $m^{x}$ and $m^{y}$ have an effect on the dispersion relations for 
the Weyl fermion so that the Weyl point is shifted to $ v_{\text{F}}^{-1}(-m^y,m^x),$
while the $z$ component of the magnetization field $m^z$ induces a spatially inhomogeneous energy gap of the Weyl fermion varying from zero to $2|m|.$
The in-plane magnetization field $m^{x,y}$ couples with the orbit of the Weyl fermion through the spin, and behave as a vector potential.  We introduce the vector potential described in terms of $m^{x,y}$ as
\begin{align} 
a^x(\boldsymbol{x},t)=-\frac{m^y(\boldsymbol{x},t)}{ev_{\text{F}}}, \quad 
a^y(\boldsymbol{x},t)=\frac{m^x(\boldsymbol{x},t)}{ev_{\text{F}}}, \label{vectorpotential}
\end{align}  
with $e<0$ the electric charge and the associated ``emergent'' electromagnetic fields
\begin{align} 
e^i(\boldsymbol{x},t)&=-\partial_t a^i(\boldsymbol{x},t),   \notag\\
b^z(\boldsymbol{x},t)&=\partial_x a^y(\boldsymbol{x},t)-\partial_y a^x(\boldsymbol{x},t), \label{electromagneticfield}
\end{align} 
with $i=x,y.$ 
As we see in the later section, 
the usage of the emergent electromagnetic fields enables us to understand clearly 
how the electric charge and current densities   are induced by the 
spatial and temporal variation of the magnetization field.

\subsection{Gradient Expansion and Keldysh Green's Function}\label{sec2gexpgreenf}
To analyze the transport phenomena of the Weyl fermion driven by the slowly-varying magnetization fields, 
we perform the Wigner transformation on the Dyson equation and make a gradient expansion.
For the details of the mathmatical treatment of the Wigner transformation and the derivation of the Keldysh 
Green's function, see Appendix \ref{appendixA}.

To construct the Keldysh Green's function we first introduce the Wigner (mixed) coordinates defined by \cite{Rammer}
\begin{align}
&X^\mu=(T, \boldsymbol{X})=\left(
\frac{t_1+t_2}{2}, \frac{\boldsymbol{x}_1+\boldsymbol{x}_2}{2}
\right), \notag\\
&x^\mu= (t, \boldsymbol{x})=\left(
t_1-t_2, \boldsymbol{x}_1-\boldsymbol{x}_2
\right),
\label{Wignercoordinates}
\end{align}
with $\mu=0,x,y.$ The coordinate $X^\mu$ is the center-of-mass coordinate describing the macroscopic characteristics of the systems, whereas
$x^\mu$ denoting the relative coordinate representing the microscopic one.
To reveal the  non-equilibrium physics driven by the magnetization fields characterized by the macroscopic variable $X^\mu$,
we derive the Fourier transform of the Green's function with respect to the microscopic variable $x^\mu,$
obtaining the Keldysh Green's function.

We perform the Wigner transformation on the non-equilibrium Dyson equation \cite{Rammer,Baraffetal,Yangetal}
\begin{align} 
&\left[K \otimes G\right] (1,2)=\hbar\delta(1-2),
\label{Dysonequation1}
\end{align} 
where $G$ represents the Green's function and $K$ the inverse of the Green's function reading
\begin{align} 
K(x_1,x_{2})=\left[
i\hbar \frac{\partial}{\partial_{t_1}}-H_0(x_1)
\right]\delta(x_1-x_{2}), 
\label{kdefinition}
\end{align} 
with the variable  $1$ or $2$ representing the combination of the space-time and spin variables, $1=(x_1,\sigma^{a,1})$ or $2=(x_2,\sigma^{a,2})$ with $a=x,y,z$,
the operator $\otimes$ denotes  the combination of the convolution with respect to  space-time and the matrix product in the spin space.
 The non-equilibrium Dyson equation  \eqref{Dysonequation1} describes the kinematics generated by magnetization fields.
 After performing the Wigner transformation on  Eq. \eqref{Dysonequation1}, we make the derivative expansion
up to the first order in $\partial_{X}$ and $\partial_{p}$ (or $\hbar$). 
The detailed derivation is presented in Appendix \ref{appendixA}.
As a result, the  first-order Green's function is given by
\begin{align} 
&\tilde{G}^{(1)}(X,p)= 
g^0 \boldsymbol{1}_{2\times2}+g^x\sigma^x+g^y\sigma^y+g^z\sigma^z, \notag\\
&g^0=\frac{
-v_{\text{F}}(\partial_{X^x}m^z)\check{p}^y+v_{\text{F}}(\partial_{X^y}m^z)\check{p}^x
-v_{\text{F}}(\partial_{X^i}m^i)m^z
}{\hbar^2\left( -E^2+
(\check{p}^x)^2+(\check{p}^y)^2+(m^z)^2
\right)^2}, \notag\\
&g^x=\frac{
-(\partial_{T}m^y)m^z+(\partial_{T}m^z)\check{p}^x
+v_{\text{F}}(\partial_{X^x}m^z)E
}{\hbar^2\left( -E^2+
(\check{p}^x)^2+(\check{p}^y)^2+(m^z)^2
\right)^2}, \notag\\
&g^y=\frac{
(\partial_{T}m^x)m^z+(\partial_{T}m^z)\check{p}^y
+v_{\text{F}}(\partial_{X^y}m^z)E
}{\hbar^2\left( -E^2+
(\check{p}^x)^2+(\check{p}^y)^2+(m^z)^2
\right)^2}, \notag\\
&g^z=-\frac{
(\partial_{T}m^x)\check{p}^x+(\partial_{T}m^y)\check{p}^y
+v_{\text{F}}(\partial_{X^i}m^i)E
}{\hbar^2\left( -E^2+
(\check{p}^x)^2+(\check{p}^y)^2+(m^z)^2
\right)^2}.
\label{firstordergreensfunction}
\end{align} 
These expressions are  the main result of this paper.
By using the above Green's function, the expectation value of the physical operators of the Weyl fermion $J=\psi^\dagger_\alpha \hat{J}_{\alpha\beta}\psi_\beta$ $(\alpha,\beta=\uparrow,\downarrow)$,
where $\psi_\alpha$ is the Weyl-fermion field operator and $\hat{J}_{\alpha\beta}$ an operator consists of Pauli matrices or the space-time derivatives, is calculated as
\begin{align} 
\langle J \rangle =-i\lim_{x \to0}
\int \frac{ dp^xdp^ydE}{(2\pi\hbar)^3}\text{Tr}\left[    \hat{J}_{\alpha\beta}     \left(     e^{-\frac{ipx}{\hbar}}\hbar^4\tilde{G}_{\alpha\beta}^{(1)}(X,p)\right)\right],
\label{physicalexpv}
\end{align}   
where $px=Et-\boldsymbol{x}\boldsymbol{p}$,
and the Tr in Eq. \eqref{physicalexpv} is taken over $2\times2$ matrices.
Since the Keldysh Green's function depends on both the coordinate $X$ and the momentum $p$, the operator  
$ \hat{J}_{\alpha\beta} $ acts not only on the plane wave $e^{-ipx/\hbar}$ but also on the Keldysh Green's function $\tilde{G}_{\alpha\beta}^{(1)}(X,p).$
Here we neglect the $X$-derivative
of the Green's function $\tilde{G}_{\alpha\beta}^{(1)}(X,p)$,
because we neglect the higher-order derivative terms of the magnetization.
We only retain the first-order derivative terms of the magnetization fields.

As we demonstrate in the following sections, 
we use the above Green's function to calculate the physical quantities such as the electric charge and current densities as well as the energy density and current.

\section{Electric Charge and Current Densities}\label{sec3electricchargecurrent}

Using the Green's functions in Eq. \eqref{firstordergreensfunction}, 
 we now study the electric charge density and current density 
induced by the magnetization fields.
We show that the electric charge as the response to the emergent magnetic field is induced 
while the Hall currents are driven as the responses to the emergent electric fields
as discussed in Ref. \cite{Nomuraetal}.

First,  the electric charge is calculated as
\begin{align} 
\rho^{(1)}_e(\boldsymbol{X},T)&=-ie\int \frac{dp^xdp^ydE}{(2\pi\hbar)^3}\text{Tr}[\hbar^4\tilde{G}^{(1)}]\notag\\
&=-\left(
\frac{e^2}{2h}\text{sgn}(m^z)
\right)b^z(\boldsymbol{X},T).
\label{chargedensity}
\end{align} 
We see that the charge density in Eq. \eqref{chargedensity} is generated as the response to the emergent  magnetic field and its sign depending of that of the $m^z$.

We next analyze the electric current density. 
The velocity operator is defined by $\hat{\boldsymbol{v}}(\boldsymbol{x})=(i\hbar)^{-1}[\boldsymbol{x},H].$
From Hamiltonian  \eqref{hamiltonian1}, we obtain  $\hat{v}^x=v_{\text{F}}\sigma^y$ and 
$\hat{v}^y=-v_{\text{F}}\sigma^x.$ Therefore,  the electric current densities are given by
\begin{align} 
j_{e}^{x,(1)}(\boldsymbol{X},T)&=-ie\int \frac{dp^xdp^ydE}{(2\pi\hbar)^3}\text{Tr}[v_{\text{F}}\sigma^y\hbar^4\tilde{G}^{(1)}]\notag\\
&=-\left(
\frac{e^2}{2h}\text{sgn}(m^z)
\right)e^y(\boldsymbol{X},T), \notag\\
j_{e}^{y,(1)}(\boldsymbol{X},T)&=-ie\int \frac{dp^xdp^ydE}{(2\pi\hbar)^3}\text{Tr}[-v_{\text{F}}\sigma^x\hbar^4\tilde{G}^{(1)}]\notag\\
&=\left(
\frac{e^2}{2h}\text{sgn}(m^z)
\right)e^x(\boldsymbol{X},T).
\label{currentdensity}
\end{align}
The currents \eqref{currentdensity} are the quantized  anomalous Hall currents generated by the emergent electric fields
with their directions depending on the sign of the $m^z.$ 
We note  that the spin density $S^{x,(1)}$$(S^{y,(1)})$ can be identified with the electric current density
$-j_{e}^{y,(1)}$$(j_{e}^{x,(1)})$$\times(ev_{\text{F}})^{-1},$ and thus, the spin densitiy can be interpreted as the response to the emergent electric field.
In contrast, the first order term in the $z$ component of the spin density $S^{z,(1)}=-i\int ( dp^xdp^ydE/(2\pi\hbar)^3)\text{Tr}(\sigma^z\hbar^4\tilde{G}^{(1)})$ is zero.

As a result, when the Weyl surface-state of the 3D TI is coupled to the magnetization field,
  the electric charge as well as electric current density are generated by the 
in-plane magnetization field  behaving as vector potential.
Then they are described as the responses to the emergent electromagnetic fields.
We note that the results in Eqs. \eqref{chargedensity} and \eqref{currentdensity}
 were also derived in the Ref. \cite{Nomuraetal} (see Eq. (5)), however, 
 their derivation is rather phenomenological assuming no sign change of $m^z.$ 
  In the present paper, we derive the results up to the first-order in the gradient expansion,
  and confirmed there appear no other terms.
 
At the end of this section, it is noted that one can easily verify from Eqs. 
\eqref{chargedensity} and \eqref{currentdensity} that the continuity equation for the electric charge
$\partial_T\rho^{(1)}_e+\partial_{X^x}j_{e}^{x,(1)}+\partial_{X^y}j_{e}^{y,(1)}=0$ is satisfied.

\section{Energy Density and Current}\label{sec4energycurrent}
In this section, we analyze the energy density and current  by using the Green's function in Eq. \eqref{firstordergreensfunction}.
We present that the energy current is  the circular current
expressed in terms of the spatial derivatives
of the $z$ component of the magnetization field, reflecting the spatial modulation of the energy gap.

To derive the energy density and current, we construct the energy-momentum tensor. 
At first from the Hamiltonian  \eqref{hamiltonian1}, the Lagrangian density in this system is
\begin{align} 
\mathcal{L}=\psi^\dagger_{\alpha}[
i\hbar\partial_0\cdot\boldsymbol{1}+i\hbar v_{\text{F}}(\sigma^y\partial_x-
\sigma^x\partial_y)-m^a\sigma^a
]_{\alpha\beta}\psi_{\beta}.
\label{lagrangiandensity}
\end{align} 
Then from the  energy-momentum tensor
$ T^{\mu\nu}=(\delta \mathcal{L}/\delta(\partial_\mu\psi)\partial^\nu\psi
+(\delta \mathcal{L}/\delta(\partial_\mu\psi^\dagger)\partial^\nu\psi^\dagger
-g^{\mu\nu}\mathcal{L}$
where $g^{\mu\nu}=\rm{diag}(1,-1,-1)$, and setting $\nu=0,$
we obtain 
\begin{align} 
T^{00}&=\psi^\dagger_{\alpha}\left[
\frac{\hbar v_{\text{F}}}{i}(
\sigma^y\partial_x-\sigma^x\partial_y)+m_a\sigma^a
\right]_{\alpha\beta}\psi_{\beta}, \notag\\
T^{x0}&=\psi^\dagger_{\alpha}\left[
i\hbar v_{\text{F}}\sigma^y\partial_0
\right]_{\alpha\beta}\psi_{\beta},\notag\\
T^{y0}&=\psi^\dagger_{\alpha}\left[
-i\hbar v_{\text{F}}\sigma^x\partial_0
\right]_{\alpha\beta}\psi_{\beta}.
\label{energymomentumtensor1}
\end{align} 
Therefore, from the Green's function \eqref{firstordergreensfunction}
the expectaion values of the energy-momentum tensors are given by
\begin{align} 
&\mathcal{E}^{(1)}=\langle T^{00}\rangle=0,\notag\\
&j_{E}^{x,(1)}=\langle T^{x0}\rangle=
-\frac{\partial_{X^y}m^z}{2\hbar}f(m^z,\Delta_{\text{b}})
, \notag\\
&j_{E}^{y,(1)}=\langle T^{y0}\rangle=
\frac{\partial_{X^x}m^z}{2\hbar}f(m^z,\Delta_{\text{b}}),
\label{energymomentumtensor2}
\end{align} 
where $\mathcal{E}^{(1)}$ denotes  the energy density,
whereas $j_{E}^{x,(1)}\left(j_{E}^{y,(1)}\right)$ the energy current for the $x(y)$ component, with
\begin{align} 
f(m_z,\Delta_{\text{b}})=\frac{1}{2\pi}\left(
\sqrt{\Delta^2_{\text{b}}+(m^z)^2}-\sqrt{(m^z)^2}
\right)
\simeq \Delta_{\text{b}}, \label{cuttofffunction}
\end{align} 
with $\Delta_{\text{b}}$ denoting the bulk gap.
Here we have introduced it to avoid the ultraviolet divergence arising from the integral 
$\int p [p^2+(m^z)^2]^{-1/2}dp$
with $p=\sqrt{(\check{p}^x)^2+(\check{p}^y)^2}$.
For instance,   $|m^z|\sim30$ meV and
$\Delta_{\text{b}}\sim0.2$ eV \cite{Leeetal}.
Thus, $|m^z|\ll \Delta_{\text{b}}$,   and the function $f(m_z,\Delta_{\text{b}})$
in Eq. \eqref{cuttofffunction} 
can be treated as a constant.
In Eq. \eqref{energymomentumtensor2}, we see  the energy current is  given by the spatial derivative of the $m^z$.
It is the term which cannot be described in terms of the emergent electromagnetic fields (or the in-plane magnetization field $m^{x,y}$) and the mechanism of its creation can be considered as the spatial variation of the energy gap $2|m^z|$. 
Furthermore, it is the current flowing circularly within the system since 
$\partial_{X^x}j_{E}^{x,(1)}+\partial_{X^y}j_{E}^{y,(1)}=0$.

\section{Conclusion}\label{sec5conclusion}
In this paper, we have investigated the charge-current densities and the energy density and current  of the 
Weyl-surface state in the 3D TI induced by the magnetization fields.  To do this, we have adopted the gradient expansion approach and derived 
the Keldysh Green's function up to the first-order derivatives with respect to the space and time.
The gradient expansion is valid when the magnetization is the spatially and temporally slowly-varying field.
The derived Keldysh Green's function describes the non-equilibium state of the Weyl fermion driven by the magnetization field, 
and enables us to study systematically the transport phenomena of the Weyl fermion.

For the electric charge density, it is described as the response to the emergent magnetic field whereas 
the electric current densities as those to the emergent electric fields with showing the quantized anomalous Hall effects. The spin densitites of the in-plane components can also be described as the responses to the emergent electric field since they are identified with the electric current densitites. 
The electric charge and current (spin) densities are induced by the in-plane magnetization fields.

The first order response of the energy density vanishes, while the energy current is  the circular current
flowing within the system.
They are represented by the spatial derivative of the $z$ component of the magnetization field. 

Our main result is the  Keldysh Green's function in Eq. \eqref{firstordergreensfunction}.
While the electric charge density, current, and spin density are represented by the emergent electromagnetic field as already discussed in Ref. \cite{Nomuraetal}, there may be other physical properties described directly by the above Green's function, e.g., 
  photon-emisison spectroscopy in non-equilibrium state.

\acknowledgements
Y.~Hama thanks Aron Beekman, Bohm-Jung Yang, and Makoto Yamaguchi  for fruitful discussion and comments.
This work was supported by RIKEN Special Postdoctoral Researcher Program (Y.~H) and 
by JSPS Grant-in-Aid for Scientific Research (No. 24224009, and No. 26103006) from MEXT, 
Japan (N.~N).

\appendix
\section{The Derivation of the Keldysh Green's Function}\label{appendixA}
We present the details of the Wigner transformation and derive the Keldysh Green's function \eqref{firstordergreensfunction}.

At first, we introduced the Wigner-transformed quantitiy which is defined by the Fourier transformation of the relative coordinate as
\begin{align} 
\tilde{A}(X,p)\equiv \int dx e^{\frac{i}{\hbar}x\cdot p}A\left(X+\frac{x}{2},X-\frac{x}{2}\right), \label{wt1}
\end{align} 
where we put the tilde for the Wigner-transformed quantity. 
Next we introduce the Wigner transformation for the convolution $C(x,y)=\int dz A(x,z)B(z,y)$ reading
\begin{align} 
\tilde{C}(X,p)&=e^{\frac{\hbar}{2i}(\partial^{\tilde{A}}_X\partial^{\tilde{B}}_p
-\partial^{\tilde{A}}_p\partial^{\tilde{B}}_X)}\tilde{A}(X,p)\tilde{B}(X,p), 
\label{wt2}
\end{align} 
where 
\begin{align} 
\partial^{\tilde{A}}_X\partial^{\tilde{B}}_p=\frac{\partial^{\tilde{A}}}{\partial T}
\frac{\partial^{\tilde{B}}}{\partial E}-\frac{\partial^{\tilde{A}}}{\partial X^i}
\frac{\partial^{\tilde{B}}}{\partial p^i} 
\end{align} 
with $i=x,y$. Here the derivatives $\partial_X^{\tilde{A}(\tilde{B})}$ or $\partial_p^{\tilde{A}(\tilde{B})}$ operates only on $\tilde{A}(\tilde{B}).$ 
The differential operator $e^{(\hbar/2i)(\partial^{\tilde{A}}_X\partial^{\tilde{B}}_p-\partial^{\tilde{A}}_p\partial^{\tilde{B}}_X)}$  in Eq. \eqref{wt2} is called the Moyal product.

We apply the Wigner transformation \eqref{wt2} to the Dyson equation \eqref{Dysonequation1} and obtain
\begin{align} 
&\left(e^{\frac{\hbar}{2i}(\partial^{\tilde{K}}_X\partial^{\tilde{G}}_p
-\partial^{\tilde{K}}_p\partial^{\tilde{G}}_X)}\right)\tilde{K}(X,p)\tilde{G}(X,p)=\hbar\cdot\boldsymbol{1}_{2\times2}, \notag\\
\label{Dysonequation2}
\end{align} 
where $\tilde{G}(X,p)$ is the Keldysh Green's function labeled by the center-of-mass coordinate $X^\mu$ 
and the conjugate momentum $p^\mu$, and
\begin{align} 
\tilde{K}(X,p)&=E\cdot\boldsymbol{1}_{2\times2}-\tilde{H}_0(X,p)=\mathcal{G}^{-1}(X,p), \notag\\
\tilde{H}_0(X,p)&=v_{\text{F}}\left(\sigma^y p^x -\sigma^x p^y\right)+{m}^a(\boldsymbol{X},T)\sigma_a.
\label{wtkandg}
\end{align} 
We expand the Moyal product in the left hand side of Eq. \eqref{Dysonequation2} up to the first order in $\partial_{X}\partial_{p}$.
Then Eq. \eqref{Dysonequation2} becomes \cite{Baraffetal,Yangetal,Freimuthetal}
\begin{align} 
\tilde{\Theta}\left[ \tilde{K},\tilde{G}      \right]\equiv
\sum_{i=0}^\infty \hbar^i \tilde{\Theta}_i\left[ \tilde{K},\tilde{G}      \right]=\hbar\cdot\boldsymbol{1}_{2\times2}.\label{thetaWT1}
\end{align} 
For instance, we have
\begin{align} 
\tilde{\Theta}_0\left[ \tilde{K},\tilde{G} \right]&=\tilde{K}\tilde{G}, \notag\\
\tilde{\Theta}_1\left[ \tilde{K},\tilde{G} \right]&=\frac{1}{2i}\left[
({\partial}_{X^\mu}\tilde{K})({\partial}_{p_\mu}\tilde{G})-({\partial}_{p^\mu}\tilde{K})({\partial}_{X_\mu}\tilde{G})
\right],\label{thetaWT2}
\end{align}
where $X_\mu=g_{\mu\nu}X_\nu$ and $p_\mu=g_{\mu\nu}X_\nu$ with $g_{\mu\nu}=\text{diag}(1,-1,-1).$
We expand the Wigner-transformed quantities  $\tilde{K}$ and $\tilde{G}$ as 
 \begin{align} 
\tilde{K}=\sum_{j=0}^\infty {\hbar}^{j} \tilde{G}^{(j)}, \quad
\tilde{G}=\sum_{j=0}^\infty {\hbar}^{j+3} \tilde{G}^{(j)}, 
\end{align} 
and substituting it to Eq. \eqref{thetaWT1}, we have
\begin{align} 
&\tilde{\Theta}_0\left[ \tilde{K},\hbar^3\tilde{G}^{(0)} \right]=\hbar^3\tilde{K}\tilde{G}^{(0)}=\hbar\cdot\boldsymbol{1}_{2\times2}, \notag\\
&\tilde{\Theta}_0\left[ \tilde{K},\hbar^4\tilde{G}^{(1)} \right]+
\hbar\tilde{\Theta}_1\left[ \tilde{K},\hbar^3\tilde{G}^{(0)} \right]
=0.
\end{align} 
As a result, we obtain the zeroth-order and the first-order Keldysh Green's functions 
\begin{align} 
 \tilde{G}^{(0)}& =\frac{1}{\hbar^2}\mathcal{G}, \notag\\
 \tilde{G}^{(1)}&=\frac{1}{2i\hbar^2}(\mathcal{G}(-\partial_{X^\mu}\mathcal{G}^{-1})\mathcal{G}(-\partial_{p_\mu}\mathcal{G}^{-1})\mathcal{G}\notag\\
&-\mathcal{G}(-\partial_{p^\mu}\mathcal{G}^{-1})\mathcal{G}(-\partial_{X_\mu}\mathcal{G}^{-1})\mathcal{G}). \label{expandgreensfunction}
\end{align} 
The expression for $\tilde{G}^{(1)}$ in terms of the magnetization field $m^a(\boldsymbol{X},T)$
and the momentum $p^\mu$ is given by Eq. \eqref{firstordergreensfunction}.

\end{document}